\documentclass[11pt]{article}
\usepackage{amssymb,amsmath,amsfonts,booktabs,geometry,ulem,graphicx,caption,color,setspace,sectsty,comment,footmisc,caption,natbib,pdflscape,subfigure,array, booktabs, ragged2e, makecell, titling}

\usepackage[hidelinks]{hyperref}
\usepackage{soul}

\usepackage[flushleft]{threeparttable}
\usepackage[linesnumbered,ruled,vlined]{algorithm2e}
\DontPrintSemicolon

\newcolumntype{L}[1]{>{\raggedright\let\newline\\arraybackslash\hspace{0pt}}m{#1}}
\newcolumntype{C}[1]{>{\centering\let\newline\\arraybackslash\hspace{0pt}}m{#1}}
\newcolumntype{R}[1]{>{\raggedleft\let\newline\\arraybackslash\hspace{0pt}}m{#1}}

\usepackage{titling}
\title{Blind targeting: Personalization under Third-Party Privacy Constraints}
\author{Anya Shchetkina \thanks{The Wharton School, University of Pennsylvania. Email: annashch@wharton.upenn.edu. I am deeply grateful to my advisor, Ron Berman, for his invaluable support in writing this paper. I also thank my committee, Ryan Dew and Raghu Iyengar, for the multitude of helpful suggestions. This paper also benefited immensely from the comments of Anna Gao, Ekaterina Goncharova, Zhenling Jiang, Nicolas Padilla, Mahdi Shahrabi, Adam Smith, Christophe Van den Bulte, and Walter Zhang.}}
\date{July 2025}

\begin{document}
\setlength{\droptitle}{-20pt}
\maketitle
\begin{abstract}
    Major advertising platforms recently increased privacy protections by limiting advertisers’ access to individual-level data. Instead of providing access to granular raw data, the platforms only allow a limited number of aggregate queries to a dataset, which is further protected by adding differentially private noise. This paper studies whether and how advertisers can design effective targeting policies within these restrictive privacy preserving data environments. To achieve this, I develop a probabilistic machine learning method based on Bayesian optimization, which facilitates dynamic data exploration. Since Bayesian optimization was designed to sample points from a function to find its maximum, it is not applicable to aggregate queries and to targeting. Therefore, I introduce two innovations: (i) integral updating of posteriors which allows to select the best regions of the data to query rather than individual points and (ii) a targeting-aware acquisition function that dynamically selects the most informative regions for the targeting task. I identify the conditions of the dataset and privacy environment that necessitate the use of such a “smart” querying strategy. I apply the strategic querying method to the Criteo AI Labs dataset for uplift modeling \citep{diemert2018large} that contains visit and conversion data from 14M users. I show that an intuitive benchmark strategy only achieves 33\% of the non-privacy-preserving targeting potential in some cases, while my strategic querying method achieves 97-101\% of that potential, and is statistically indistinguishable from Causal Forest \citep{athey2019generalized}: a state-of-the-art non-privacy-preserving machine learning targeting method.
\end{abstract}

\section{Introduction}
Over the last decade, questions of privacy and data protection have become central to legislative discussions and marketers' decision-making. In 2018, the General Data Protection Regulation (GDPR) came into effect in the European Union. And while in the United States there is still no federal statute regulating online data privacy, several states have enacted GDPR-like laws. In 2020, California became the first state to pass a comprehensive consumer data-protection law, the California Consumer Privacy Act (CCPA). By July 2025, 18 other states had passed similar statutes, and bills were in committee in 10 more.\footnote{https://iapp.org/resources/article/us-state-privacy-legislation-tracker/. Accessed on June 30, 2025.}

In response to the current and in anticipation of future consumer data regulations, many companies have rolled out various versions of privacy protection mechanisms. For instance, to learn emoji popularity, Apple distorts the true emoji usage at a user's device before collecting and aggregating the data \citep{apple2017dp}. Meta is experimenting with limiting access to offsite cookies \citep{wernerfelt2025estimating}, and on the government side, the US Census releases demographic tables protected by differential privacy \citep{census2023dpbrief}.

A particularly important class of protections arises when data must be shared across organizations (e.g., between advertisers and advertising platforms). Instead of raw individual-level data, the platform provides a privacy-preserving query interface (often an SQL workspace) that advertisers can only use to learn aggregate statistics of the data (e.g., ``average conversion rate of females under age 30 in July 2025''). To further protect individual users, the number of such aggregate queries is limited, and noise is added to the results. A notable example of such an interface is Google's Privacy Sandbox and Ads Data Hub \citep{google2024adh}. Other examples include LinkedIn \citep{linkedin2019ppa}, as well as AWS Clean Rooms \citep{aws2024cleanroomsdp} and Snowflake \citep{snowflake2024dp}, who develop and provide these interfaces for other companies to use.

Accessing the data through limited and noisy aggregate queries does not necessarily pose a challenge for all marketing problems. For example, estimating an overall result of an A/B test, i.e., computing the average treatment effects can be done easily with privacy protections. However, more sophisticated analyses become difficult to perform. In particular, if an advertiser aims to design a targeting policy based on the results of an experiment, third-party privacy protection interfaces render standard tools for estimating heterogeneous treatment effects \citep[e.g., Causal Forest,][]{athey2019generalized} unusable, as these tools rely on access to individual-level data.

This paper asks if targeting and personalization are still possible under such privacy protections, and how much of the targeting value advertisers can retain. To answer this question, I develop a new targeting method, called strategic querying, that preserves privacy protections by only accessing a limited amount of aggregate data, yet allows marketers to dynamically extract the information most useful for targeting. I compare this method with an intuitive privacy-preserving benchmark, uniform querying, which divides the entire data range into equal-sized bins and queries each of them, under a variety of data-generating processes and levels of privacy protections.

The strategic querying method is based on the principles of Bayesian Optimization. In contrast to uniform querying, it dynamically updates its beliefs and uncertainty about the function it is trying to learn to select the next best query. A Bayesian optimization approach offers several important advantages over a simple uniform method. First, it allows one to exploit the sequential nature of querying and thus incorporate the information from the previous queries into future ones. Second, it is goal-informed, and thus guides the data exploration away from uninformative wasteful queries, which is particularly important when the number of queries is strictly limited. Third, it enables marketers to seamlessly integrate their prior knowledge about the data and past experiments via priors.

However, the classic Bayesian Optimization framework cannot be applied as is for targeting based on aggregate queries. I therefore extend the framework and introduce two innovations. First, Bayesian Optimization is designed to sample \textit{points} of a function. In contrast, in privacy-protected data exploration interfaces, a marketer is required to run aggregate queries, thereby selecting \textit{regions} of the data to query. I build on the results of \cite{smith2018gaussian} to extend the posterior predictive distributions to regions, and thus make it possible to select the next best region to query (this is called integral updating).

Second, Bayesian Optimization is typically used to maximize an objective function, such as finding the best combination of hyperparameters for a neural network. In contrast, when designing a targeting policy based on estimated heterogeneous treatment effects, marketers do not look for the maximum of HTEs, and instead aim to select the appropriate action for each combination of covariates. Therefore, the rules that guide goal-driven data exploration in Bayesian optimization are motivated by an incorrect goal for targeting and may not result in an optimal exploration sequence. To align the data exploration with the targeting goal, I introduce a novel targeting-aware acquisition function that looks for regions of the data that are the most consequential for a targeting policy. This rule allows marketers not to waste limited queries on learning about regions that will not affect the end goal. For example, if a region clearly has a positive treatment effect and thus should be targeted, it makes sense not to spend a query on this region and instead query the one in which the sign of the treatment effect is unknown. The combination of Bayesian Optimization with integral updating and the targeting-aware acquisition function constitutes the strategic querying method.

To test the strategic querying method, I conduct a series of simulation studies. I find that both the targeting-aware acquisition function and control over the size of queried regions are necessary for robust performance of the method across various data-generating processes. Across simulation settings, the strategic querying method is able to uncover 67\% of the oracle non-privacy-protected targeting value (the difference between the optimal personalized policy under full information and the best blanket intervention). The 33\% gap is likely an overestimation because in reality, even in non-privacy-preserving settings, the oracle value is hard to achieve when estimating heterogeneous treatment effects because of irreducible noise in the data \citep[see, e.g.,]{shchetkina2024heterogeneity}. Simulation results show that the strategic querying method, as well as other privacy-preserving methods in general, suffer most when: (1) treatment effects vary a lot across people with similar characteristics, and (2) the number of allowed queries is small. Both of these factors, however, are less likely in realistic conditions, which gives reason to expect solid performance from the strategic querying method.

Finally, in an empirical application, I study the effectiveness of the strategic querying method in a real-world setting. I use the Criteo AI Labs dataset for uplift modeling \citep{diemert2018large} (an anonymized compilation of A/B tests with treatment and visit variables of 14M users with 12 covariates) to compare the performance of a non-privacy-preserving state-of-the-art machine learning targeting method \citep[Causal Forest,][]{athey2017digital} with two privacy-preserving querying methods: the strategic querying method and a simpler uniform querying benchmark that splits the data into equal-sized ranges for each covariate. I impose four levels of privacy protections on the dataset. I find that the uniform querying method is unstable, dropping to 33\% of the Causal Forest performance in one of the settings. In contrast, the strategic querying method reliably attains 97--101\% of the Causal Forest performance across all settings.

The analysis of the empirical application suggests several insights. First, in contrast to the uniform querying method, where the granularity of queries depends on the query limit, strategic querying adjusts query sizes dynamically, starting with broad questions first and then narrowing it down to regions of interest. Second, information sharing that occurs in the strategic querying method due to smoothing priors allows the method to distinguish effectively between the inherent noise in the data and the added differential privacy noise, mitigating the impact of privacy protections compared to uniform querying. Third, the fact that the strategic querying method performed on par with a state-of-the-art machine learning method, while relying on fewer than 50 noisy values instead of 600,000 raw data entries, highlights that targeting is an information-light task when segments are broad enough, and thus is not adversarial to privacy goals. 

These results provide an opportunity for marketers to feasibly target even under restrictive third-party privacy protections. The effectiveness of the targeting policy would, however, depend on the ability of marketers to specify the concrete downstream policy goals and to align the data exploration pipeline accordingly.

The remainder of the paper is organized as follows. Section \ref{sec:background} provides background on privacy, targeting, and methodological foundations of strategic querying. Section \ref{sec:setting} formalizes the mathematical notation and setting. Section \ref{sec:methods} introduces the strategic querying method, as well as an intuitive privacy-preserving benchmark, called uniform querying. Section \ref{sec:simulation} describes the simulation study and compares the performance of the strategic querying method against the benchmark under a variety of data-generating processes. Section \ref{sec:application} provides an empirical application. Section \ref{sec:conclusion} concludes.

\section{Related Literature} \label{sec:background}

This paper is related to four streams of literature. Substantively, it is linked to work on privacy and on personalization. Methodologically, it relates to applications of Bayesian optimization and learning from aggregate data.

Concurrent with the growth of privacy concerns in society \citep{goldfarb2012shifts}, the marketing literature has also seen an increase in privacy-related research. The main areas of study are substantive in nature and include understanding and measuring consumers' valuations of privacy \citep[e.g.,][]{athey2017digital, lin2022valuing, lin2025choice}, adoption of privacy-enhancing technologies by consumers and firms \citep[e.g.,][]{johnson2020consumer, johnson2024unearthing}, as well as consequences of privacy regulations \citep[see ][for a review]{dube2024intended}. This stream of literature showed that privacy regulations lead to an array of adverse effects on the effectiveness of online advertising \citep{goldfarb2011privacy, tucker2014social, wernerfelt2025estimating, gu2025can}, on market concentration \citep{campbell2015privacy, johnson2023privacy}, on innovation \citep{bleier2020consumer}, and on polarization \citep{bondi2023privacy}.

In contrast, the computer science literature on privacy has been primarily theoretical or methodological. Early results showed an inherent trade-off between preserving privacy and data utility \citep{dinur2003revealing}. If the data is not perturbed properly or enough (thereby losing some useful information), the original private values can be reconstructed \citep[e.g.,][]{cohen2018linear, garfinkel2019understanding}. To quantify the risk of such reconstruction, \cite{dwork2006differential} introduced the concept of differential privacy (DP), which since then has become fundamental in theoretical research and implementations of privacy protections. Differential privacy can be achieved in various ways, usually called mechanisms. A common mechanism is adding noise to the summary statistic of interest (such as the average of a variable). The scale of the noise determines the level of protection against re-identification of private values. However, in practice, choosing the level of protection, and therefore, the level of noise, is not a straightforward task \citep{hsu2014differential, dwork2019differential}. In addition, existing algorithms have been modified to satisfy DP requirements. Examples include privacy-preserving classification trees \citep{friedman2010data}, deep learning \citep{abadi2016deep}, construction of synthetic data \citep{rosenblatt2022spending}, and best arm identification in multi-armed bandits \citep{chen2024fixed}. These papers consider ``first-party" privacy protection: a researcher can only observe privacy-protected results, yet the algorithm has access to the true underlying data. In contrast, this paper focuses on a setting with third-party privacy protection: an unperturbed dataset belongs to an advertising platform, which offers a privacy-protected interface for a researcher to interact with, so that any algorithm that a researcher uses never sees the original data.

The second stream of relevant literature relates to personalization and targeting. With the availability of vast data on consumer characteristics and behavior, personalization of interventions has gained potential to greatly enhance revenue and customer retention, and thus has received much attention in the marketing literature \citep[see, e.g.,][]{rafieian2021targeting, yoganarasimhan2023design, dube2023personalized}. Most recent methodological advances in this space have focused on ways to improve the study design pipeline prior to running a personalization algorithm itself. Examples include using information from previous campaigns to improve targeting \citep{ibragimov2020improving, huang2024incrementality}, sequential experimentation \citep{chen2024policy}, and optimal sample size calculation \citep{simester2025sample}. Another area of interest is non-standard settings, such as long-term targeting \citep{huang2024doing}, coarse \citep{zhang2024coarse}, or multiobjective \citep{rafieian2025multiobjective} personalization. For all these applications of personalization, having access to granular and accurate individual-level data is essential. Privacy protections, therefore, naturally come into conflict with the effectiveness or even possibility of personalization.

Two papers stand out as closest to the present work, studying the interplay between privacy and personalization. \cite{korganbekova2023balancing} document that personalized recommendations on Wayfair suffer from browsing history clearing, which results in user fragmentation. They show that probabilistic matching of users' browsing histories can mitigate this problem. \cite{huang2023debiasing} study the estimation of conditional average treatment effects (CATEs) under local differential privacy, where each entry of the dataset is perturbed. They demonstrate that in this setting, such perturbation acts as measurement noise and propose a post-processing approach to debias CATEs. In comparison, this paper studies personalization when a granular dataset is not observed at all and can only be interacted with via a restricted interface (such as Google or AWS Privacy Sandbox).

Methodologically, this paper adds to a growing literature on applications of Bayesian optimization (BO). Bayesian optimization is a method of strategically sampling the objective function to get close to its maximum in as few steps as possible \citep{kushner1964new, garnett2023bayesian}. It is usually applied in scenarios when the objective function is hard or expensive to evaluate, such as when tuning hyperparameters in machine learning models \citep{snoek2012practical}. Recently, these ideas have been brought to adaptive experimental design \citep{greenhill2020bayesian, rainforth2024modern}. In marketing, \cite{dew2024adaptive} used the BO framework to adaptively elicit consumer preferences over images. I contribute to this literature by developing a custom acquisition function that is suited for targeting rather than maximization. In addition, this paper combines the BO idea of strategic sampling with the idea of learning from aggregate data \citep[e.g.,][]{musalem2008s, xu2025multitask}. Most relevant in this stream is \cite{smith2018gaussian} who show how Gaussian Processes can be used to nonparametrically estimate a latent function when only its averages over ranges are observed. Drawing on these results, I extend the Bayesian optimization framework to sampling averages of ranges instead of points (as in the classic setting).

\section{Setting and notation} \label{sec:setting}

This paper focuses on the following setting of third-party privacy protections.\footnote{At the time of writing (2025), this is a stylized version of privacy protections employed by Google Ads and AWS Clean Rooms. Similar SQL workspaces for data querying are also commonly employed by other companies \citep{shi2024scalable}.} In this setting, an advertising platform is running an experiment and stores the data, while a marketer can only access aggregate-level results. 

A marketer is running an A/B test on an advertising platform.\footnote{Random assignment is not necessary but simplifies the exposition by abstracting away from endogeneity problems.} The platform collects data on consumers and the performance of ads. The dataset $\mathcal{D}$ consists of consumer characteristics $X_i$, treatment assignment $W_i \in \{0, 1\}$, and outcome $Y_i$. The goal of the marketer is to infer a targeting policy $\pi: X \to A$, where $A \in \{0,1\}$ is the chosen treatment for a given set of consumer characteristics. An effective targeting policy would maximize
\begin{equation} \label{targ_objective}
    \mathbb{E} \Bigl[ (Y_i^1 - c)\mathbb{I}\bigl(\pi(X_i) = 1\bigr) + Y_i^0\mathbb{I}\bigl(\pi(X_i) = 0\bigr) \Bigr]
\end{equation}
where $Y_i^1, Y_i^0$ are the potential outcomes of consumer $i$ after being exposed to treatment and control, respectively, $c$ is the cost of treatment, and $\mathbb{I}$ is the indicator function.

Traditionally, a marketer would use an individual-level dataset $\mathcal{D}$ to estimate a flexible machine learning model that would predict outcomes or treatment effects on the individual level \citep[e.g., a causal forest by][]{athey2019generalized} and assign each person to the condition with the highest outcome (or equivalently, a positive treatment effect). 

\begin{figure}
    \centering
        \caption{Querying example} 
    \includegraphics[width=\linewidth]{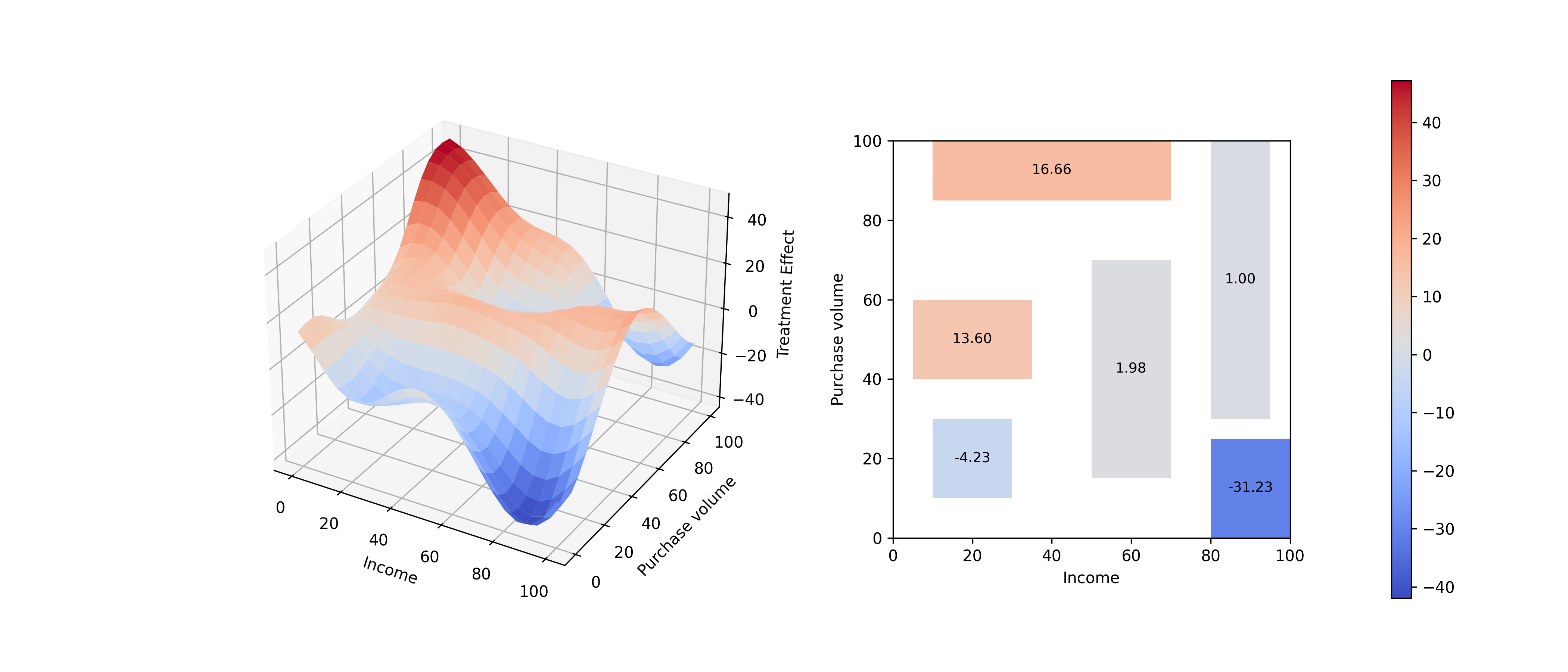}
    
    \footnotesize \justifying On the left: a hypothetical treatment effect surface over income and purchasing volume. On the right: the result of 6 queries to the dataset on the left. Each rectangle represents $[\underline{X_q}, \overline{X_q}]$, and the color value of each rectangle is the result of query $q$ (CATE in this rectangle).
    \label{fig:query_ex}
\end{figure}

However, the advertising platform employs third-party privacy protections and does not allow the marketer to access the individual-level dataset $\mathcal{D}$. Instead, a marketer can make a series of $Q$ queries to this dataset through an interface provided by the platform. The result of each query $q$ is the difference in averages of the outcomes $Y$ over the chosen range of consumer characteristics $X$ under treatment and control plus zero-mean noise $\eta_q$, which will be explained below:
\begin{equation} q = 
    \frac{\sum_{i} \Bigl[Y_i \mathbb{I}\bigl(X_i \in [\underline{X_q}, \overline{X_q}], W_i = 1\bigr)\Bigr]}{\sum_{i} \Bigl[ \mathbb{I}\bigl(X_i \in [\underline{X_q}, \overline{X_q}], W_i = 1\bigr)\Bigr]} -  \frac{ \sum_{i} \Bigl[Y_i \mathbb{I}\bigl(X_i \in [\underline{X_q}, \overline{X_q}], W_i = 0\bigr)\Bigr]}{\sum_{i} \Bigl[\mathbb{I}\bigl(X_i \in [\underline{X_q}, \overline{X_q}], W_i = 0\bigr)\Bigr]} + \eta_q
\end{equation}
where $[\underline{X_q}, \overline{X_q}]$ is a range of user characteristics (a hyperrectangle) that a marketer is interested in sampling at query $q$. Figure \ref{fig:query_ex} shows an example of such querying.

Under random assignment (and the standard assumptions of ignorability and overlap), the expected output of such a query is equal to:
\begin{equation} \mathbb{E}(q) = 
    \mathbb{E}(Y_i^1 - Y_i^0|X_i \in [\underline{X_q}, \overline{X_q}])
\end{equation}
To further protect consumers against re-identification, the platform adds noise $\eta_q \sim N(0, \sigma_q)$ to the outputs of the query (called a Gaussian mechanism in the differential privacy literature). The scale of noise is inversely proportional to the number of people affected by the query: 
\begin{equation} \label{eq:dp_noise}
    \sigma_q = \frac{s}{\sum_{i}\bigl[ \mathbb{I}(X_i \in [\underline{X_q}, \overline{X_q}])\bigr]}
\end{equation}
Given this privacy framework, the goal of the marketer is to conduct a series of queries $q = \{1,\dots, Q\} $ to the dataset, estimate conditional average treatment effects, and design a targeting policy $\pi$ that would optimize the targeting objective (Equation \ref{targ_objective}). To achieve this, the marketer must solve two challenges that are not present in a standard targeting setting with individual-level data: (1) the data is only available at the aggregate level in the form of group averages, and (2) the number of such averages is limited and thus the marketer must be strategic in choosing the queries in order to maximize the amount of information. These two challenges make applying established targeting algorithms, such as causal forests \citep{athey2019generalized} or policy trees \citep{zhou2023offline}, impossible.

\section{Querying methods} \label{sec:methods}
In this section, I introduce two strategies that a marketer can use to choose a sequence of queries, estimate the CATEs, and design a targeting policy. The first one is uniform querying: an intuitive and non-adaptive strategy, which I will use as a benchmark. The second is the focus of this paper, strategic querying: an adaptive policy-aware sampling strategy based on Bayesian optimization. 

\subsection{Uniform querying}
\begin{figure}
    \centering
    \caption{Uniform querying} 
    \includegraphics[width=\linewidth]{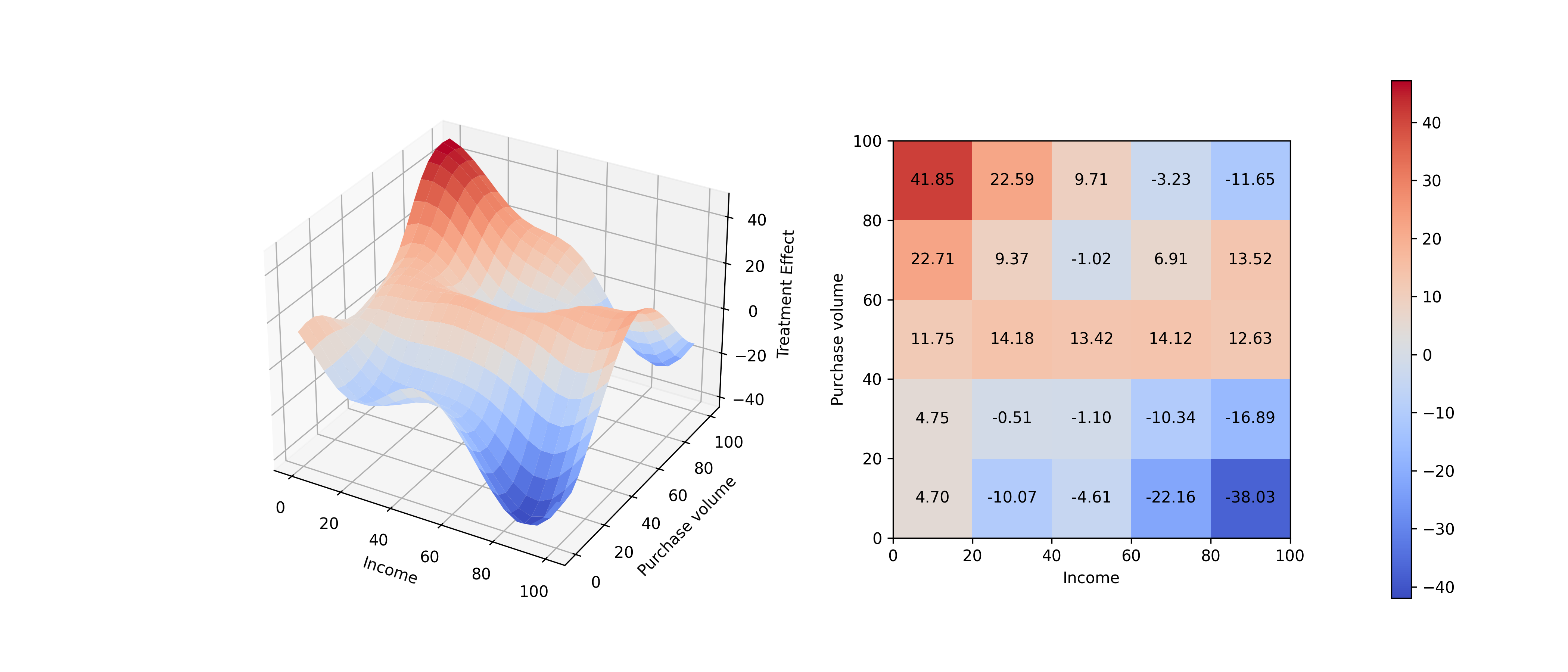}
    
    \footnotesize \justifying On the left: a hypothetical treatment effect surface over income and purchasing volume. On the right: the result of 25 uniform queries to the dataset on the left. Each rectangle represents $[\underline{X_q}, \overline{X_q}]$, and the color value of each rectangle is the result of query $q$ (CATE in this rectangle). A targeting policy (given $c = 0$) is to assign all positive cells to treatment and all negative to control. 
    \label{fig:query_uniform}
\end{figure}

One intuitive approach that a marketer can use to query the database is to divide all variables into equal-sized ranges, so that the entire range of $X$ is divided into cells of the same size.\footnote{Of course, this is possible only if the number of queries $Q = R^V$, where $R$ is the number of bins along each variable, and $V$ is the number of variables. In other cases, a uniform strategy would be to make the cells as equal in size as possible.} For each of the resulting cells, the marketer can query the database and receive the CATE in the cell. A targeting policy following such querying is also straightforward: assign all cells whose query output is higher than the cost $c$ to treatment and assign the remaining cells to control. An example of such querying with two variables and 25 queries (so that each variable is divided into 5 equal bins) is shown in Figure \ref{fig:query_uniform}.

This querying approach has two potential disadvantages. First, as Figure \ref{fig:query_uniform} shows, uniform querying does not use the limited queries efficiently because it is not aware of the task the results will be used for. For instance, the cell in the bottom right corner is surrounded by highly negative cells, so most likely, it is also negative and should not be treated. Thus, it might be better to forego querying this cell and instead spend the query on trying to divide the cell between $(20,20)$ and $(40,40)$, which is located between a positive and a negative cell, to find out which portion of it should be treated.

\begin{figure}
    \centering
    \caption{A problem with uniform querying} 
    \includegraphics[width=0.7\linewidth]{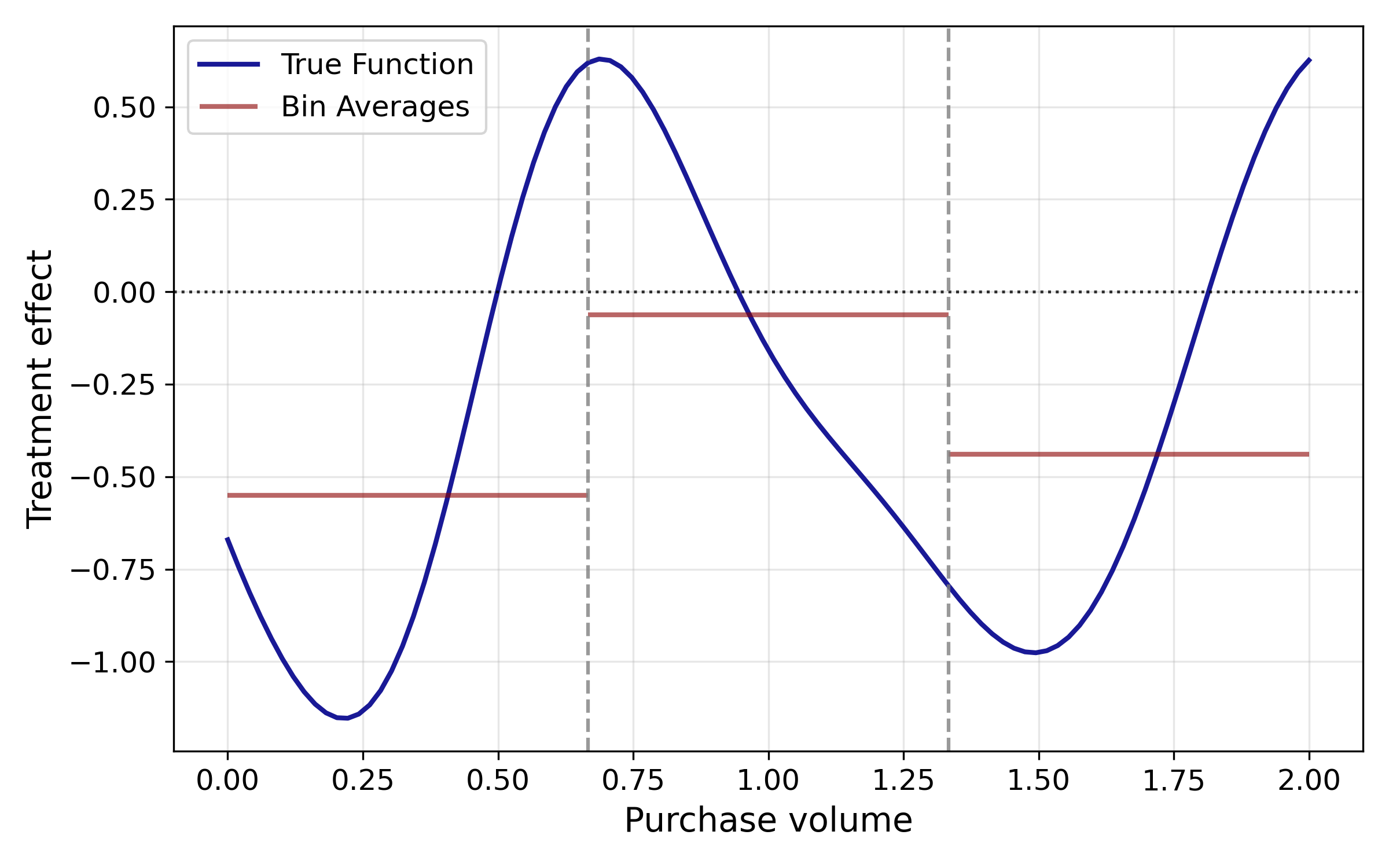}
    
    \footnotesize \justifying The blue line represents treatment effects for different values of purchase volume. An optimal targeting policy would target regions above zero, however, uniform querying with 3 bins (results are represented by horizontal red lines) misses all of these regions and assigns everyone to control.
    \label{fig:uniform_problem}
\end{figure}

The second disadvantage is that the bins are rigid and predetermined, and thus may not be able to capture the relevant customer segments. This case is illustrated in Figure \ref{fig:uniform_problem}. The figure shows a hypothetical treatment effect curve over purchase volume. If a marketer has access to this curve (i.e., is able to observe individual-level data), they can design an efficient targeting policy that assigns customers with a positive treatment effect (in case of a zero cost) to treatment and those with a negative treatment effect to control. On Figure \ref{fig:uniform_problem}, this corresponds to targeting customers with purchasing volumes between 0.5 and 0.95 and above 1.8. However, if a marketer does not have access to individual-level data and uses the uniform approach to query the dataset (with 3 bins in this example), the results of all three queries would be negative. This would lead the marketer to erroneously believe that the entire population has a negative treatment effect and should not be treated.

Next, I introduce a novel approach, called strategic querying, that improves on the shortcomings of the uniform approach. First, it is policy-aware, and thus uses the limited queries efficiently on the regions that are the most relevant for the targeting task. Second, it is dynamic and adaptive, and thus can capture the ranges of variables which are relevant for customer segmentation.

\subsection{Strategic querying} Strategic querying, aimed at addressing shortcomings of uniform querying, is based on the ideas of Bayesian optimization, which makes it adaptive. However, the standard Bayesian optimization framework cannot be mapped directly into the current setup. Therefore, I develop two extensions to the Bayesian optimization framework: (1) sampling of ranges instead of points and (2) a targeting-aware acquisition function. Sampling of ranges instead of points allows the method to be applied in aggregate data settings, and the targeting-aware acquisition function increases the efficiency of limited queries. In this subsection, I first introduce the general framework of Bayesian optimization, and then describe the two extensions.

\subsubsection{Bayesian optimization}
Bayesian optimization \citep{kushner1964new} is an adaptive sampling method to optimize functions that are hard or expensive to evaluate. Examples include tuning hyperparameters in machine learning models \citep{snoek2012practical}, reaction optimization in synthetic chemistry \citep{shields2021bayesian}, and drug discovery \citep{pyzer2018bayesian}. The general idea of the method is to replace the hard-to-evaluate objective function with a proxy function that is easy to evaluate and, importantly, to quantify the uncertainty of. Then, using the estimation of the expected value and its uncertainty for the proxy function, a researcher can choose the next point to sample by balancing between points with high expected value (exploitation) and high uncertainty (exploration). 

\begin{figure}
    \centering
    \caption{Bayesian optimization example} 
    \includegraphics[width=0.7\linewidth]{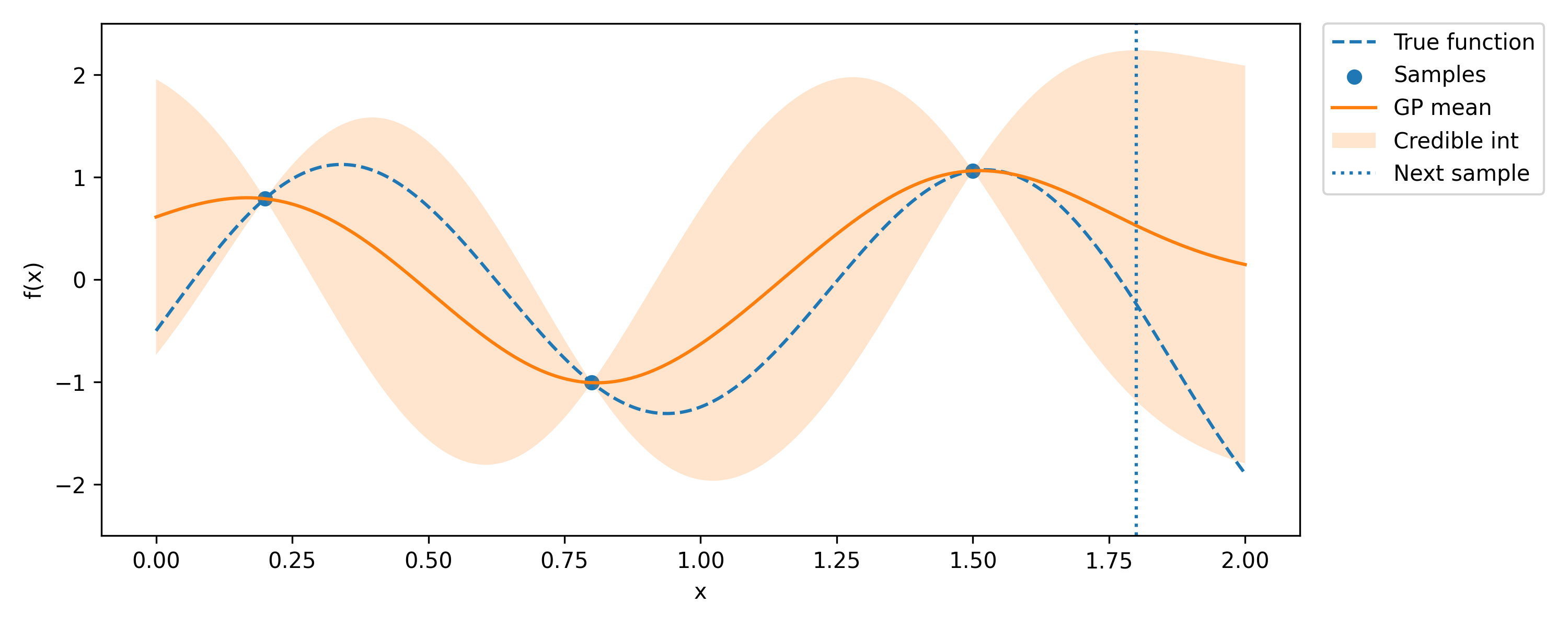}
    
    \footnotesize \justifying A hard-to-evaluate objective function (blue dashed line) is estimated with a GP proxy function (orange solid line). The expected value and its uncertainty for the proxy function are then used to construct an acquisition function that balances the two (upper bound of the shaded region). The next point to sample is selected as the maximum of the acquisition function (vertical dotted line).
    \label{fig:bo_example}
\end{figure}

Figure \ref{fig:bo_example} provides an illustration of Bayesian optimization. In this figure, the objective function (black dashed line) is expensive to evaluate, so a researcher started with three samples (red points). Now, the researcher wants to select the next (fourth) point to sample to get as close to the maximum of the objective function as possible. To do this, the researcher first fits a proxy function (blue line with a shaded area) through the sampled points. A proxy function that is most widely used in the literature and applications is a Gaussian Process (GP).\footnote{Gaussian Processes are also widely used outside of the context of Bayesian optimization. For recent examples see \cite{dew2024adaptive} and \cite{dew2024your}.} A Gaussian Process, $f(x) \sim \mathcal{GP} (m(x), k(x,x'))$, is a stochastic process in which any collection of points is jointly normally distributed. It is defined by its mean function, which captures the expected value of the function, $m(x)$ and a kernel $k(x,x')$ which captures the covariance between the values of the function at different points. GPs offer two important advantages in the context of Bayesian optimization. First, they are nonparametric, and thus can capture the unknown objective function. Second, being a collection of normal distributions, they provide natural and closed-form uncertainty quantification.

After fitting a GP to the three sampled points and deriving the expected value and the uncertainty for all candidate points for sampling, the researcher uses an acquisition function that balances expectation and uncertainty to select the next point to sample. A common example of an acquisition function is the upper confidence bound (UCB): $UCB (x) = \mu(x) + 1.96\sigma(x)$. Such a function is easy to maximize (as opposed to the original objective), and the next point to sample is selected as the maximum of the acquisition function (vertical dotted line on the figure).

\begin{figure}
    \centering
    \caption{Extensions of Bayesian optimization framework} 
    \includegraphics[width=0.45\linewidth]{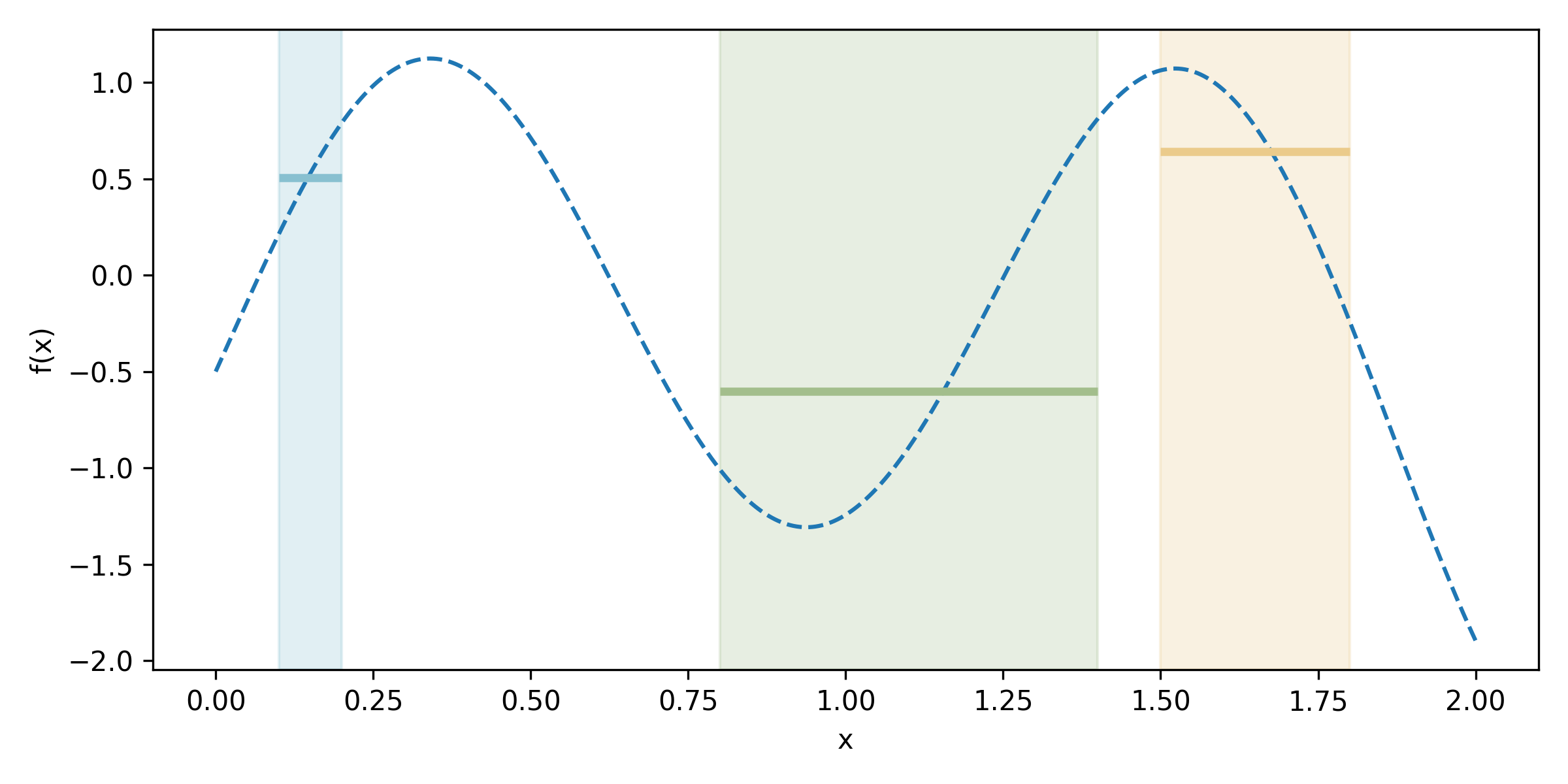}
    \includegraphics[width=0.45\linewidth]{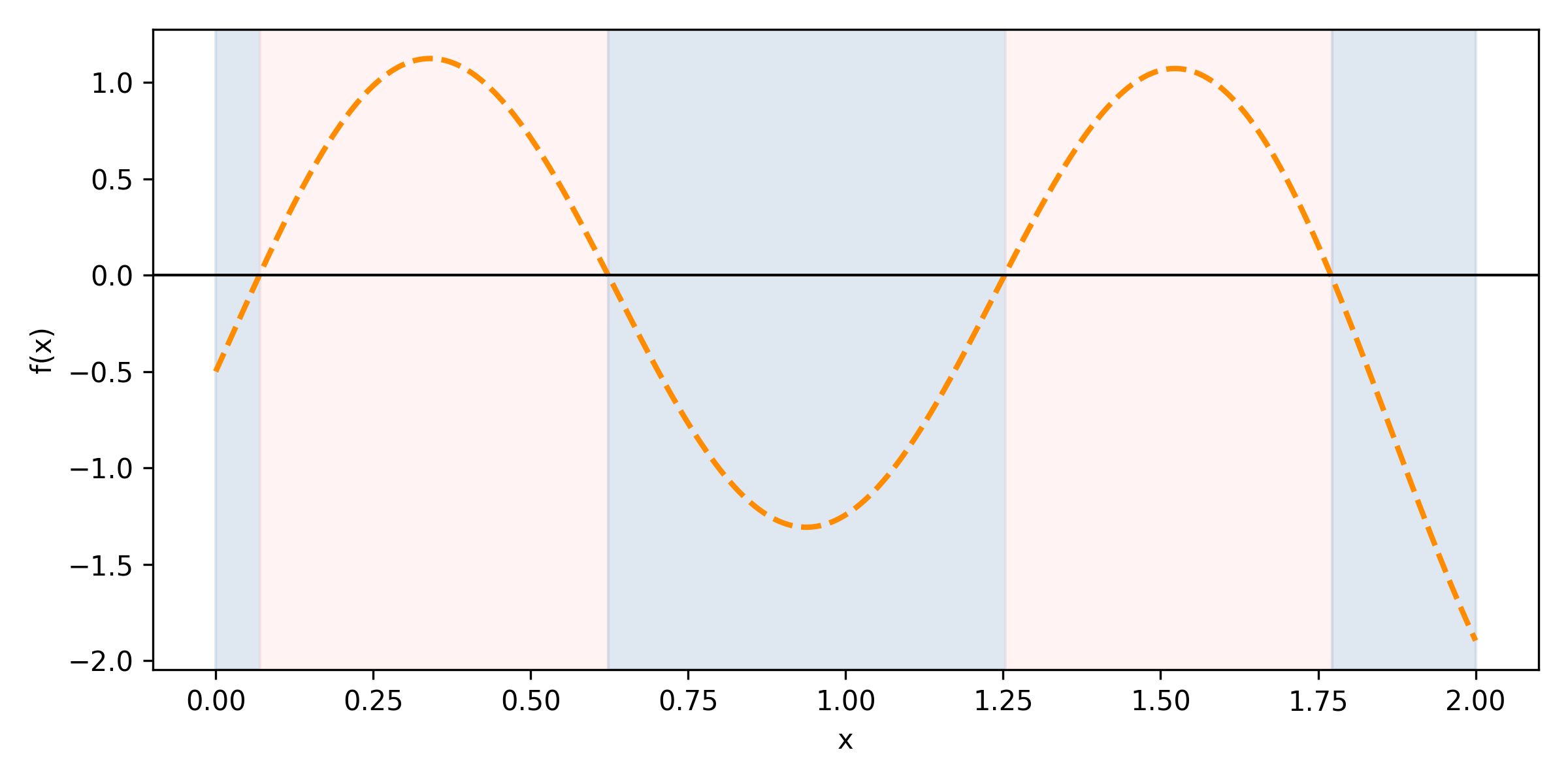}
    
    \footnotesize \justifying On the left: a marketer samples and learns from averages over ranges $[\underline{X_q}, \overline{X_q}]$ instead of points. On the right: the goal is to find regions where the treatment effects are positive or negative and not to find the maximum.
    \label{fig:extensions}
\end{figure}

As the figure shows, this algorithm allows to approach the maximum of the objective function within a small number of iterations, which makes it particularly valuable when the objective function cannot be sampled many times, either because it is computationally expensive or, as in case of privacy protections, there is a limit on the number of queries. However, this algorithm cannot be directly applied to the targeting setting in Section \ref{sec:setting}. First, in this setting a marketer must query regions of varying sizes $[\underline{X_q}, \overline{X_q}]$ instead of points as in the standard Bayesian optimization framework (left panel of Figure \ref{fig:extensions}). Second, the goal of the algorithm is not to find the maximum of the treatment effect function but instead to determine for which $X$ it is positive and for which it is negative. Next, I extend the standard Bayesian optimization framework to incorporate these two differences.

\subsubsection{Sampling regions instead of points}
As stated above, the first challenge in applying the Bayesian optimization framework to the setting in Section \ref{sec:setting} is that a marketer needs to sample averages over ranges of $X$ instead of points. For the ideas of Bayesian optimization to be applicable, we need to know the expected value and its uncertainty for such an average over each candidate region. Gaussian Processes are closed under linear functionals \citep{williams2006gaussian}, and averages over a range can be thought of as an integral which is a linear functional. In particular, for a squared exponential kernel (one of the most widely used in the literature), the expressions for the mean and the variance of an integral can be derived in closed form. \cite{smith2018gaussian} use the following version of the squared exponential kernel:\footnote{It differs from the standard formula by $\sqrt{2}$ for ease of computation.}
\begin{equation} \label{eq:SE_kernel}
 k(x, x') = \alpha e^{-\frac{(x-x')^2}{l^2}}   
\end{equation}
This kernel is characterized by two parameters: amplitude $\sqrt{\alpha}$, which controls the magnitude of the function, and lengthscale $l$, which controls how fast the function can change. Figure \ref{fig:gp_example} shows examples of a squared-exponential GP with different values of $\sqrt{\alpha}$ and $l$. The effects of these parameters are illustrated in Figure \ref{fig:gp_example}. \cite{smith2018gaussian} derive the following kernel for averages over a range:
\begin{figure}
    \centering
    \caption{Gaussian processes example} 
    \includegraphics[width=0.7\linewidth]{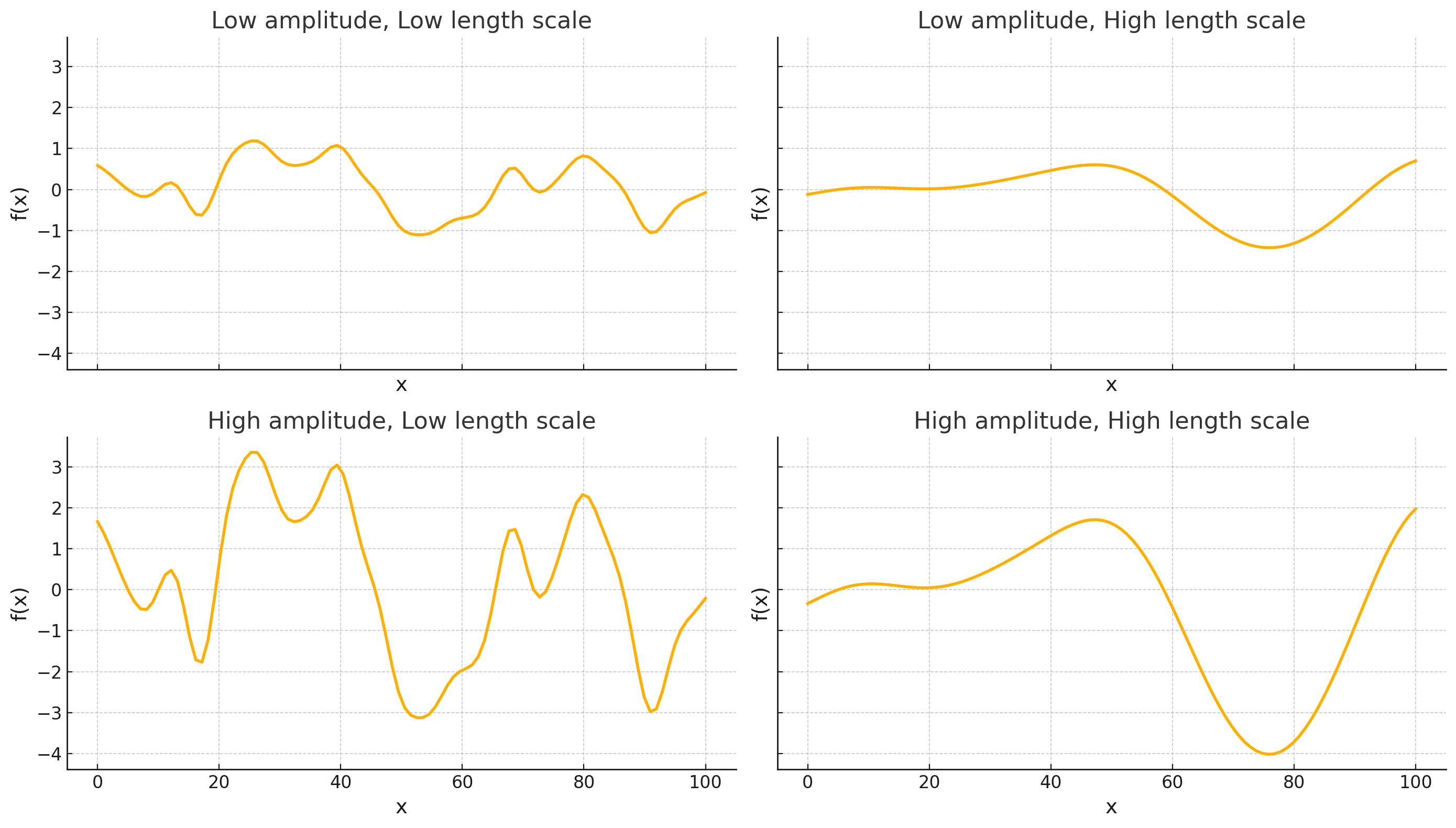}
    
    \footnotesize \justifying The effects of changing GP hyperparameters: amplitude $\sqrt{\alpha}$ and lengthscale $l$ in Equation \ref{eq:SE_kernel}.
    \label{fig:gp_example}
\end{figure}
\begin{equation} \label{eq:integral_kernel}
k_{FF}((s,t), (s',t')) = \alpha \frac{l^2}{2} \left[ g\left(\frac{t - s'}{l}\right) + g\left(\frac{t' - s}{l}\right) - g\left(\frac{t - t'}{l}\right) - g\left(\frac{s - s'}{l}\right) \right]
\end{equation}
Here, $(s, t)$ and $(s', t')$ are ranges over which the averages are computed, and $g(x) = x\sqrt{\pi} \mathrm{erf}(x) + e^{-x^2}$.

I further combine this result with the formula for the posterior predictive of noisy observations \citep{williams2006gaussian}:
\begin{align}
m_* &= \mathbf{k}_*^\top (K + \sigma^2 I)^{-1} \mathbf{y} \\
V_* &= k(\mathbf{x}_*, \mathbf{x}_*) - \mathbf{k}_*^\top (K + \sigma^2 I)^{-1} \mathbf{k}_*
\end{align}
where $m_*$ and $V_*$ are the posterior predictive mean and variance for an average of a function over a given range, $\mathbf{k}_*$ is the vector of covariances between a given range and observed ranges according to Equation \ref{eq:integral_kernel}, $K$ is the matrix of such covariances between observed ranges, $\mathbf{y}$ is a vector of observed averages, and $\sigma^2$ is a scale of noise (e.g., the Gaussian mechanism in differential privacy). 

These combined results allow us to compute the expectation of an average over a range and its variance for a proxy GP function, which in turn allows to define an acquisition function over any range $[\underline{X_q}, \overline{X_q}]$ and perform Bayesian optimization over ranges. Next, I describe an acquisition function which is appropriate for a targeting task.

\subsubsection{Policy-aware acquisition function}

The second difference between the classic Bayesian optimization framework and the targeting setting of this paper (Section \ref{sec:setting}) is that in the former, the goal is to maximize the function we sample from (Figure \ref{fig:bo_example}), while in the latter, the goal is to find a targeting policy based on the function we sample from (right panel of Figure \ref{fig:extensions}). Thus, acquisition functions such as UCB that look for high expected value are not well-fitted for the task. Instead, I introduce an alternative targeting-aware acquisition function (TAAF)\footnote{Relatedly, \cite{misra2019dynamic} modify UCB in multi-armed bandits to incorporate economic theory of demand.}:
\begin{equation} \label{eq:taaf}
    TAAF([\underline{X_q}, \overline{X_q}]) = \beta V([\underline{X_q}, \overline{X_q}]) - |m([\underline{X_q}, \overline{X_q}])|
\end{equation}
where $[\underline{X_q}, \overline{X_q}]$ is a candidate range for sampling, and $m([\underline{X_q}, \overline{X_q}])$ and $V([\underline{X_q}, \overline{X_q}])$ are the posterior predictive mean and variance over this range.

\begin{figure}
    \centering
    \caption{Intuition for acquisition function} 
    \includegraphics[width=0.7\linewidth]{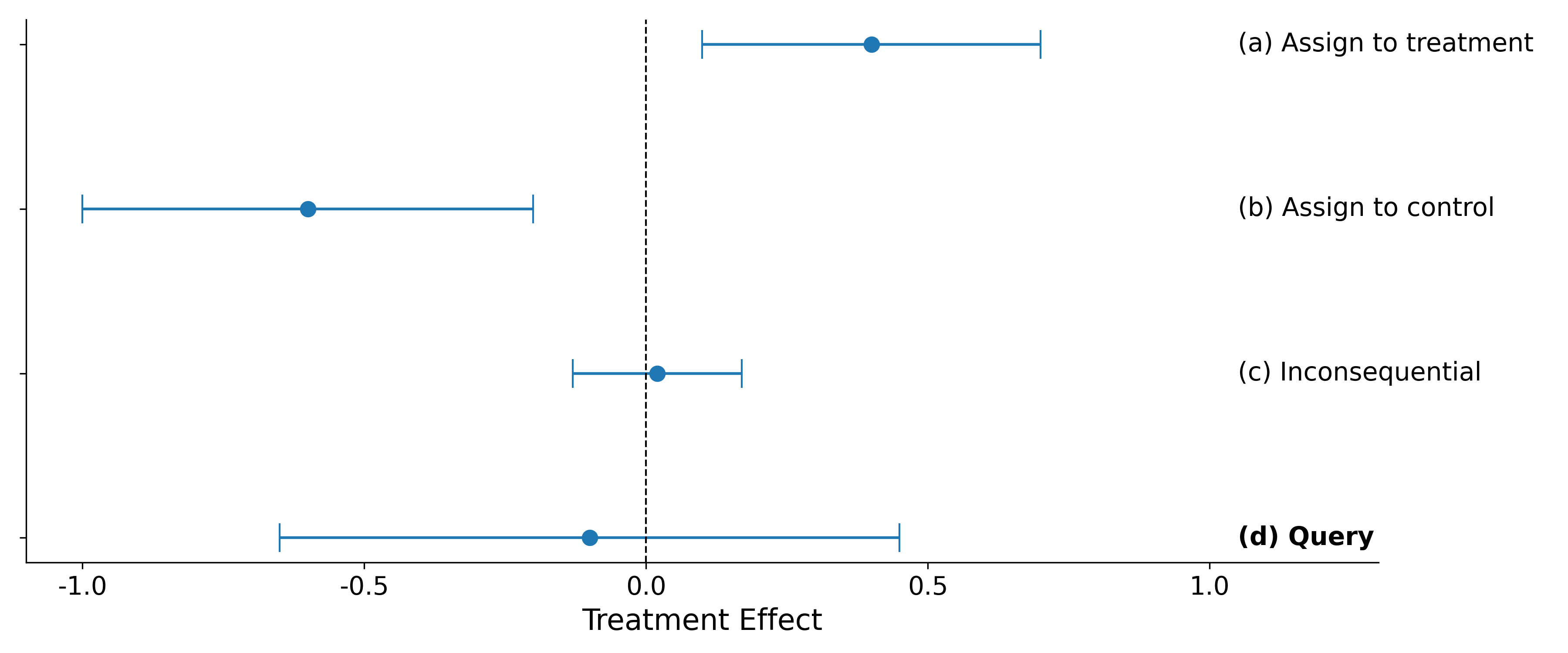}
    
    \footnotesize \justifying Candidate regions to query: (a) high expected value relative to variance, should be assigned to treatment regardless; (b) low expected value relative to variance, should be assigned to control regardless; (c) expected value is close to zero and variance is small, assignment does not matter much; (d) expected value is close to zero and variance is large, querying is needed.
    \label{fig:af_intuition}
\end{figure}

This function looks for regions with high uncertainty but close to zero expectation.\footnote{For the cost of treatment $c>0$, this cost should be subtracted from the sampled average values.} The intuition behind it is displayed in Figure \ref{fig:af_intuition}. If a region's expected value is very high (a) or very low (b), it should either be assigned to treatment or control, and the value of additional information is low.\footnote{This property of a targeting task was also noted in, for example, \cite{zhao2012estimating}, \cite{zhang2023optimal}, and \cite{chen2024fixed}.} Therefore, for a queried region to provide meaningful information, the expected value of this region should be close to zero. However, if the uncertainty around this close-to-zero value is small (c), assignment to either treatment or control is inconsequential, because effect sizes are small regardless. An ideal region to query thus would have an expected value close to zero but high uncertainty (d).

Mathematically, this can be seen by performing an analysis similar to \cite{shchetkina2024heterogeneity}. Suppose that a unit's treatment effect $\tau$ is distributed normally with mean $\mu$ and variance $\sigma^2$. For simplicity, I will assume that we can query this unit and reveal its true treatment effect, after which we make a decision on whether to target it or not. The expected value of querying can be expressed as the expected value of a targeting policy after revealing the true treatment effect of a unit minus the expected value of a targeting policy prior to this. The former is the expectation of a rectified normal distribution and the latter is the maximum between the outcome under treatment ($\mu$) and control ($0$):\footnote{Again, the cost of treatment $c$ is assumed to be equal to zero for simplicity, otherwise, it should be deducted.}
\begin{equation}
    \mu \left[1 - \Phi\left(-\frac{\mu}{\sigma}\right)\right] + \sigma\phi\left( - \frac{\mu}{\sigma}\right) - \max\{\mu, 0\}
\end{equation}
Differentiating this with respect to $\mu$ when $\mu > 0$:
\begin{equation*}
    - \left[\Phi\left(-\frac{\mu}{\sigma}\right) + \mu \phi\left(-\frac{\mu}{\sigma}\right)\left(-\frac{1}{\sigma}\right)\right] - \sigma \left[\phi\left(-\frac{\mu}{\sigma}\right)\left(-\frac{\mu}{\sigma}\right)\left(-\frac{1}{\sigma}\right)\right] = - \Phi\left(-\frac{\mu}{\sigma}\right) <0
\end{equation*}
Similarly, when $\mu < 0$:
\begin{equation*}
    1 - \left[\Phi\left(-\frac{\mu}{\sigma}\right) + \mu \phi\left(-\frac{\mu}{\sigma}\right)\left(-\frac{1}{\sigma}\right)\right] - \sigma \left[\phi\left(-\frac{\mu}{\sigma}\right)\left(-\frac{\mu}{\sigma}\right)\left(-\frac{1}{\sigma}\right)\right] = 1 - \Phi\left(-\frac{\mu}{\sigma}\right) > 0
\end{equation*}
Therefore, the maximum expected value of querying is achieved when the expectation of the treatment effect $\mu$ is zero.

Now, taking the derivative with respect to $\sigma$ for any $\mu$,
\begin{equation*}
     -\mu\phi\left(-\frac{\mu}{\sigma}\right)\left(\frac{\mu}{\sigma^2}\right) + \left[\phi\left(-\frac{\mu}{\sigma}\right)-\sigma\phi\left(-\frac{\mu}{\sigma}\right)\left(-\frac{\mu}{\sigma}\right)\left(\frac{\mu}{\sigma^2}\right)\right] = \phi\left(-\frac{\mu}{\sigma}\right) > 0
\end{equation*}
Therefore, the maximum expected value of querying is increasing with uncertainty $\sigma$. Taken together, these results motivate an acquisition function that prioritizes regions with high variance and close to zero expectation, such as TAAF (Equation \ref{eq:taaf}).

The discussion abstracted away from the size of the queried regions, the fact that the queried regions provide information about adjacent regions, and the limit on the number of queries, which makes this a dynamic programming problem. Therefore, TAAF should be considered an approximation heuristic for selecting regions, rather than a fully optimal solution.

\section{Simulation study} \label{sec:simulation}
I conduct a simulation study to compare the two querying methods (uniform and strategic) and determine how characteristics of the data-generating process and the privacy environment influence their performance. Furthermore, I analyze the performance of alternative acquisition functions for strategic querying and incorporate consideration of the sizes of queried regions.

\subsection{Setting}
\textbf{Data-Generating Process.} For all simulation exercises, I model the treatment effects as a Gaussian Process over three variables ranging from 0 to 100 with a zero mean function and a squared exponential kernel.\footnote{An example of a Gaussian Process over two variables is shown in Figure \ref{fig:query_ex}.} I vary the amplitude and the lengthscale parameters of the GP (see Figure \ref{fig:gp_example}) among several levels. The amplitude takes values of 2 (lowest magnitude of treatment effects), 5, and 10 (highest magnitude of treatment effects), while the lengthscale takes values of 10 (treatment effects change a lot between customers with similar covariates), 30, and 50 (treatment effects change little between customers with similar covariates).

\textbf{Privacy Environment.} To model different levels of privacy protections, I vary the limit on the number of queries $Q$ and the scale of differential privacy noise $s$ added to the results of queries.\footnote{The scale of noise added to a query $q$ depends on $s$ and on the number of people affected by this query as in Equation \ref{eq:dp_noise}.} The number of queries ranges between 8 (lowest), 27, 64, and 125 (highest). I chose a cubic number to allow for an easy implementation of a uniform querying benchmark: the limits correspond to dividing every variable into 2, 3, 4, and 5 regions correspondingly. The scale of the differential privacy noise takes values of 0.1 (least privacy-protected), 1, 10, and 100 (most privacy-protected).

\textbf{Acquisition Functions.} Alongside TAAF (Equation \ref{eq:taaf}), I test two additional acquisition functions for strategic querying.\footnote{These functions were chosen from a larger set of candidate functions after a pretest, see details in the Appendix.} Both functions prioritize regions with high variance, as is customary in Bayesian optimization. However, the first one, $AF_{var}$, is purely variance-based and reflects mutual information about the sign for regions that are close to zero in expectation:
\begin{equation} \label{eq:af_var}
    AF_{var} = \sigma^{2} - \frac{\sigma^{2}}{\sigma^{2} + \tau}\,\ln\!\left(1 + \frac{\tau}{\sigma^{2}}\right)
\end{equation}
The second function, $AF_{regret}$, approximates the regret from misspecifying the sign for a region (and thus handles $|\mu|$ in the opposite way to TAAF).
\begin{equation} \label{eq:af_regret}
     AF_{regret} = \sigma|\mu|
\end{equation}

\textbf{Size of Regions.} Finally, neither of the heuristic acquisition functions explicitly incorporates the sizes of regions queried. Theoretically, considering the sizes of regions to query might be beneficial: regions that are too large naturally have a high variance, however, they might not provide meaningful information about granular subgroups. On the other hand, regions that are too small might use limited queries inefficiently by not learning about the entire population, while also being affected more by differential privacy noise. To test if the size of the region matters, I construct four versions of each acquisition function:
\begin{enumerate}
    \item Original AF, size not explicitly considered.
    \item An added penalty for larger regions
    \item Candidate regions restricted to mid-sized
    \item Candidate regions restricted to mid-sized and an added penalty for larger regions.
\end{enumerate}

In total, the simulation study consists of 144 parameter settings ($3 \times 3$ for the data-generating process and $4 \times 4$ for the privacy environment). For each parameter setting, I ran 100 repeats and for each repeat computed (a) the value of an oracle targeting policy, (b) the value of a targeting policy constructed using uniform querying, (c) 12 values of the targeting policies constructed using strategic querying (three acquisition functions each with four versions of region size considerations).

\subsection{Results}

\begin{figure}
    \centering
    \caption{Strategic vs uniform querying across all settings} 
    \includegraphics[width=\linewidth]{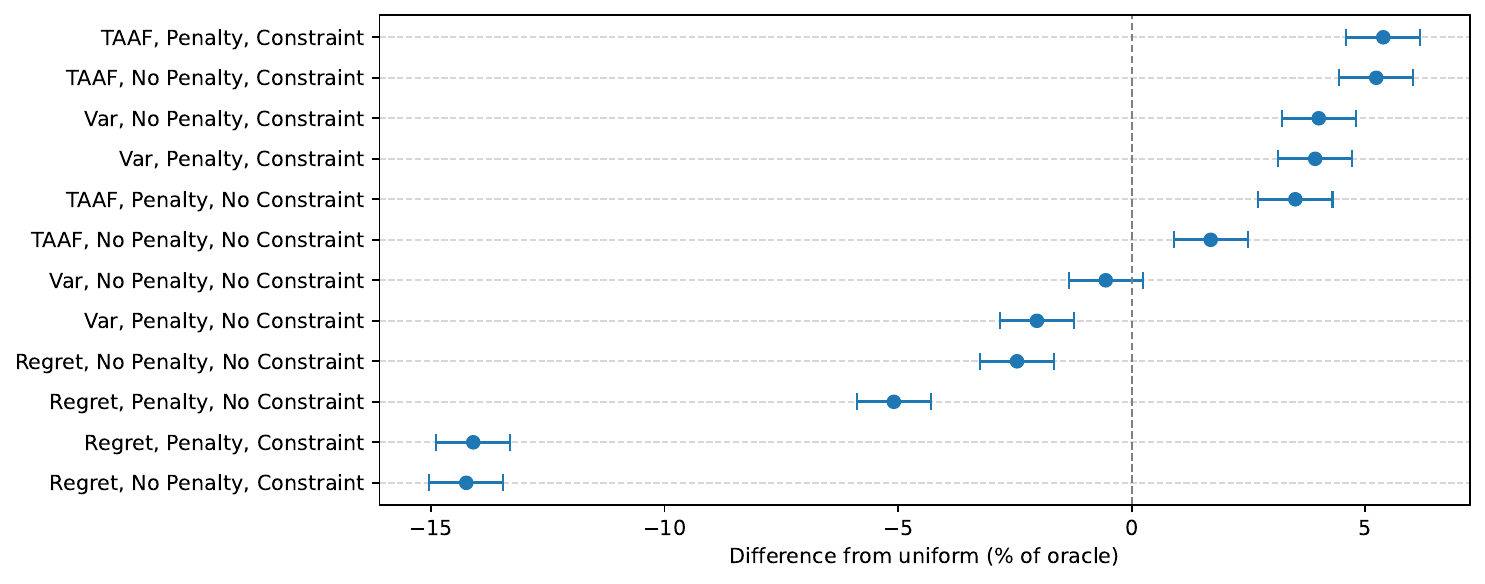}
    
    \footnotesize \justifying The aggregate performance of strategic querying against uniform querying with different acquisition functions across all 144 simulation settings and 100 runs for each. The performance is scaled by the value of the oracle targeting policy. 
    \label{fig:sim_overall}
\end{figure}

\textbf{Acquisition Functions.} First, I compare the performance of 12 Bayesian optimization models against uniform querying across different simulation settings. Figure \ref{fig:sim_overall} shows the results of the simulation study pooled across all 144 settings and 100 repeats. The difference between strategic and uniform querying is scaled by the value of the oracle targeting policy. There are several insights from this analysis. First, the choice of the acquisition function matters for strategic querying. The regret function performs the worst, staying even below uniform querying. On the other hand, all versions of TAAF perform better than uniform querying on average. Second, it is important to consider the sizes of regions. For both TAAF and the variance acquisition functions, the versions that constrain the candidate regions perform best, with the difference being particularly large for the variance-only acquisition function. An acquisition function that relies purely on variance will have a baked-in tendency to favor large regions and might not uncover relevant and granular CATEs, which is why size constraints are essential for its good performance.

\begin{figure}
    \centering
    \caption{Pairwise method comparison} 
    \includegraphics[width=0.9\linewidth]{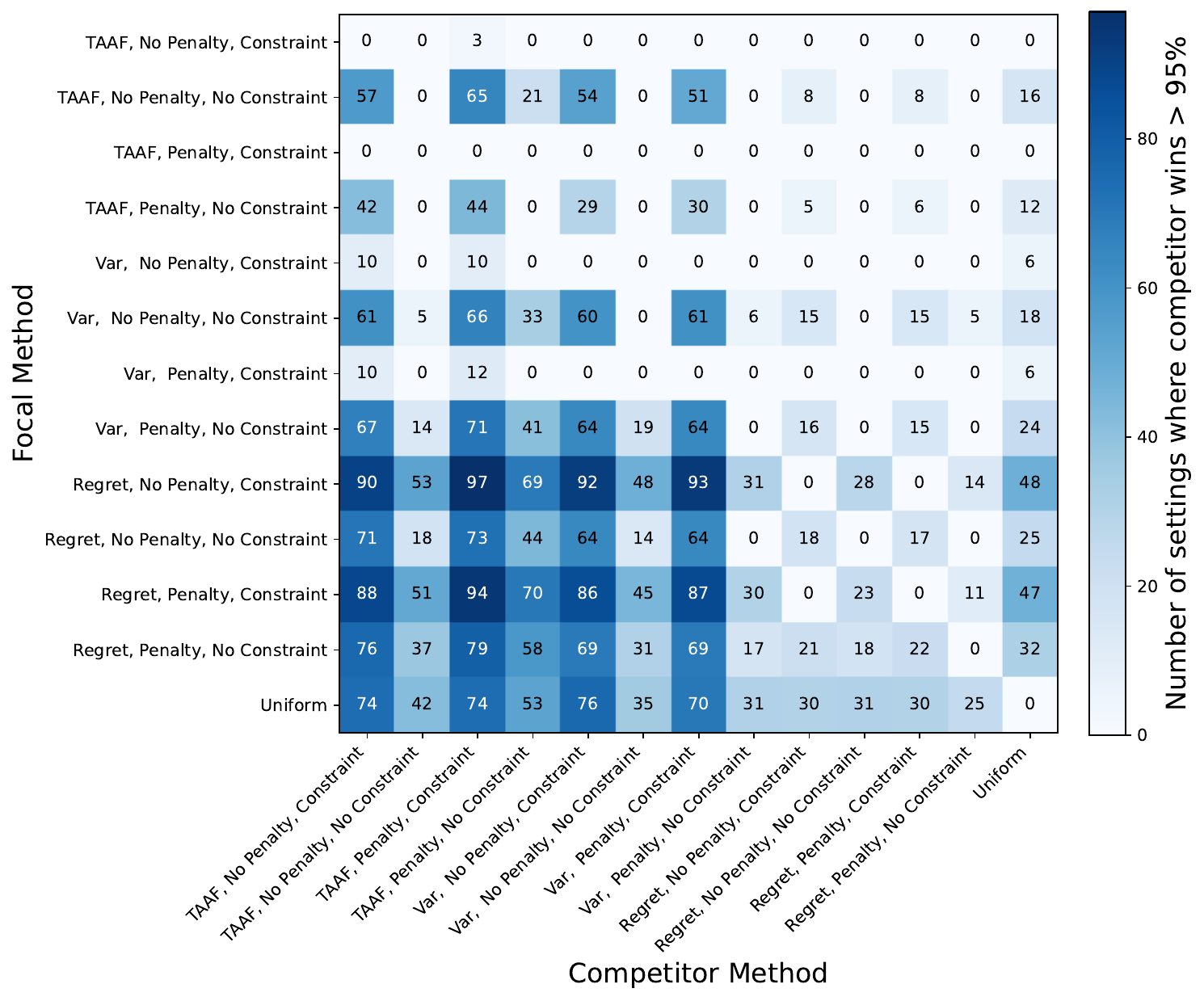}
    
    \footnotesize \justifying For each focal method (rows) and each competitor method (columns), the value of a cell represents the number of simulation settings (out of 144) when a competitor method performs better in at least 95\% of runs (i.e., the focal method is dominated by the competitor method in this setting).
    \label{fig:sim_dominance}
\end{figure}

However, even if some methods do well on average, there might still be particular DGP or privacy environment settings where they do not do as well relative to other methods. To explore this possibility, I conduct a pairwise setting-level dominance analysis, which is presented in Figure \ref{fig:sim_dominance}. This figure shows a $13\times13$ (12 strategic and 1 uniform querying methods) matrix, in which rows represent a focal method, and columns --- a competitor method. The values denote the number of simulation settings (out of 144), in which the competitor method outperformed the focal method in at least $95\%$ of repeats, in other words, the number of settings where a focal method is dominated by a competitor method. This analysis demonstrates that the only method that is \textit{not dominated by} any other method in any setting is TAAF with a size penalty and a size constraint. In addition, this method \textit{dominates} every other method in at least some settings. TAAF with no size penalty but with a size constraint is a close second: it is only dominated by the best method in 3 settings but is not dominated by anything else in any setting, and dominates every method except for the best one in some settings. Taken together, these results show that TAAF with constraints on candidate regions not only perform best on average but also demonstrate robustly superior performance across a range of DGP parameters and privacy environments. 

\begin{figure}
    \centering
    \caption{Strategic vs uniform querying: performance by parameter levels} 
    \includegraphics[width=0.9\linewidth]{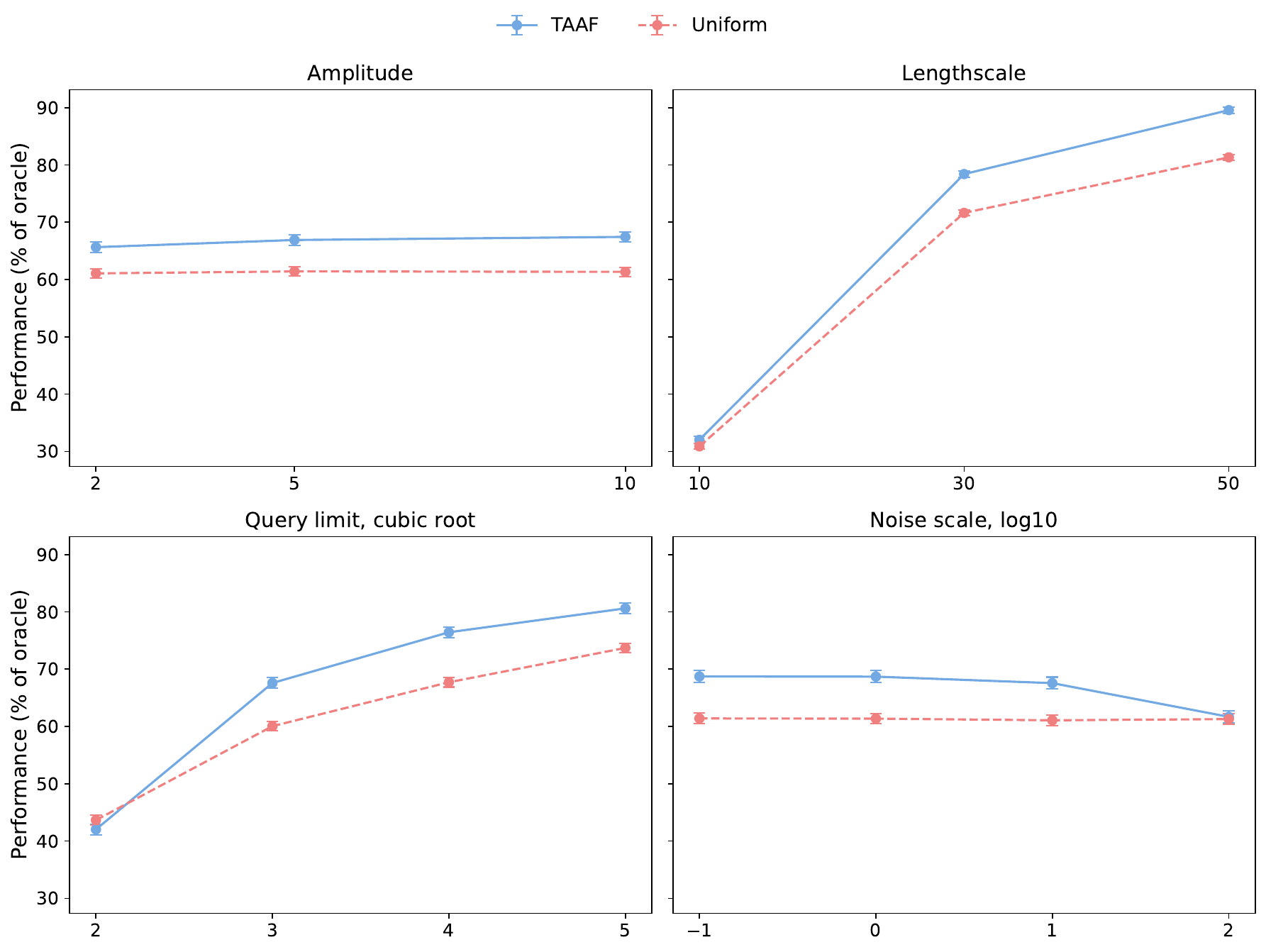}
    
    \footnotesize \justifying The average performance and the confidence interval (in percentages of the value of the oracle policy) for the best strategic querying method (TAAF, size penalty, constrained) and the uniform querying method for different levels of parameters in the simulation settings. E.g., for the top left plot, the performance across all settings and repeats with DGP amplitude set to 2 is aggregated (same with 5 and 10).
    \label{fig:sim_params}
\end{figure}

\textbf{Effects of DGP parameters.} Henceforth, I will focus on the comparison between uniform querying and the best strategic querying (TAAF, with size penalty and constraint), which are displayed in Figure \ref{fig:sim_params}.

\textbf{Amplitude.} Overall, the magnitude of treatment effects did not have a major effect on the performance of the uniform and the strategic querying methods (see top left panel of Figure \ref{fig:sim_params}). In particular, uniform querying was not affected by the magnitude of treatment effects at all, while strategic querying delivered a slightly better targeting value when treatment effects were larger rather than smaller, while retaining a robustly better performance over the uniform method at all levels of amplitude.

This is due to the fact that, holding the noise level constant, a higher amplitude of the treatment effects increases the signal-to-noise ratio for every queried region. In this simulation, the signal-to-noise ratio impacts the strategic querying method more than it does the uniform. For the uniform querying method, the signal-to-noise ratio can only result in an incorrect assignment of a region to treatment or control if the magnitude of noise is so high relative to the magnitude of the treatment effect that it flips the sign of the average over the queried region. On the other hand, the strategic querying method is impacted through the same channel and an additional one. Regions with low signal-to-noise ratio make the posterior predictive distribution less accurate, and thus lead strategic querying to select suboptimal regions for querying. These simulation results, however, were derived under the assumption that the only source of noise in the data is the differential privacy noise added to the results, driving signal-to-noise baseline ratio quite high, which results in uniform querying not being impacted by that variation. In realistic scenarios, however, the CATEs are often inherently noisy, and the signal-to-noise ratio is low, so the variation in it would most likely matter for uniform querying as well.

\textbf{Lengthscale.} In contrast to the amplitude of treatment effects, their lengthscale (how much treatment effects oscillate along covariates) greatly affects the performance of both uniform and strategic querying methods (see top right panel of Figure \ref{fig:sim_params}). In particular, for the smallest setting of lengthscale (treatment effects change most rapidly for close values of covariates), both methods deliver similarly poor performance. This is not surprising: if the treatment effect surface is very ``jagged'', relevant customer segments for targeting are very small and thus are unlikely to be captured by a limited number of group averages. However, as long as the lengthscale is not very small, further gains from increasing it are modest. In realistic datasets, we expect the treatment effects to change relatively little along covariates, and thus for privacy-preserving methods and strategic querying in particular to perform well.

\textbf{Number of queries.} Similar to lengthscale, the number of queries allowed has the biggest impact when its value is extreme. For a limit of 8 queries, both strategic and uniform methods perform poorly, with the uniform method slightly but insignificantly outperforming the strategic method (see 
bottom left panel of Figure \ref{fig:sim_params}). This is due to the fact that the Bayesian optimization algorithm does not have enough time to properly calibrate the posterior. Once the number of queries increases, however, the Bayesian optimization algorithm starts to robustly outperform the uniform querying method. Interestingly, the difference between the two methods has an inverted-U shape. For a small number of queries, both methods fail to capture heterogeneity, while for a large number of queries, a simple method such as uniform querying is granular enough to uncover relevant segments. It is for the medium numbers of queries that the advantage of the strategic querying becomes the most apparent: it has enough time to learn something, while choosing what to learn optimally.

\textbf{Noise scale.} Like amplitude, the noise scale also affects signal-to-noise ratios. The mechanisms are the same: for the uniform benchmark, it only affects the chances of incorrect estimation of the sign of a cell, while for the strategic querying, it also impacts the posterior distribution at all steps and thus leads to suboptimal querying. Regardless of these adverse effects, strategic querying significantly outperforms the uniform benchmark for all levels of noise except for the highest one (see bottom right panel of Figure \ref{fig:sim_params}). Similar to the amplitude effects, these relative disadvantages of strategic querying for high levels of noise might not translate to real datasets where the data-generating process is noisy itself.

Overall, the results of the simulation study show that (1) the choice of the acquisition function for the strategic querying method is critical to achieve stable performance, (2) restrictions and penalties on the size of queried regions can further improve the results, and (3) strategic querying is expected to outperform a simpler uniform querying benchmark for a variety of realistic data-generating processes.

\section{Empirical Application} \label{sec:application}
\subsection{Data}
To test the performance of strategic querying in realistic scenarios, I use the uplift modeling benchmark dataset published by Criteo AI Labs \citep{diemert2018large}. This dataset was constructed by assembling and resampling the results of several incrementality tests and consists of almost 14M rows, each with 12 anonymized user features, a treatment indicator (84.6\% were treated), as well as visit and conversion indicators. The summary statistics of user features are displayed in Table \ref{tab:criteo_summary}.

\begin{table}[]
    \centering
    \caption{Summary statistics of user features in Criteo Uplift dataset}
    \begin{tabular}{lcccccccccccc}
\toprule
{} &        f0 &        f1 &       f2 &       f3 &       f4 &       f5 &       f6 &       f7 &       f8 &       f9 &      f10 &      f11 \\
\midrule
Mean &   19.62 &   10.07 & 8.45 &  4.18 & 10.34 &  4.03 &  -4.16 & 5.10 &  3.93 & 16.03 &  5.33 & -0.17 \\
SD   &   5.38 &   0.10 &  0.30 &  1.34 & 0.34 &  0.43 &  4.58 &  1.21 &  0.06 &  7.02 &  0.17 &  0.02 \\
25\% &  12.62 &  10.06 & 8.21 & 4.68 & 10.28 & 4.12 & -6.70 & 4.83 & 3.91 & 13.19 & 5.30 & -0.17 \\
50\% &   21.92 &   10.06 & 8.21 &  4.68 & 10.28 &  4.12 &  -2.41 & 4.83 &  3.97 & 13.19 &  5.30 & -0.17 \\
75\% &   24.44 &   10.06 &  8.72 &  4.68 &  10.28 &  4.12 &  0.29 &  4.83 &  3.97 &  13.19 &  5.30 &  -0.17 \\
\bottomrule
\end{tabular}
    \label{tab:criteo_summary}
\end{table}
Since in this dataset ``treatment'' corresponds to ad eligibility and ``control'' users do not see ads, I add a cost of the treatment. The data is anonymized and rescaled, so I assume the cost of treatment to be $c = 0.01$ (visit is recorded as $1$, and no visit is $0$). Such operationalization of the cost of treatment creates meaningful heterogeneity across customers (some should be assigned to treatment and some to control) and thus a non-trivial optimal targeting policy.

To quantify variation in the performance of querying methods and targeting policies, I take advantage of the size of the Criteo dataset. I bootstrap 100 samples of 50,000 customers each for estimation of targeting policies, and similarly 100 samples of 50,000 customers each for their Inverse Propensity Weighting (IPW) evaluation. IPW is a standard approach for estimating the value of counterfactual policies from experimental data that is free from winner's curse resulting from errors in estimating heterogeneous treatment effects \citep[see, e.g.,]{simester2020targeting, rafieian2023ai, andrews2024inference}:
\begin{equation} \label{eq:ipw}
    \widehat{IPW}(\pi) = \frac{1}{n} \sum_{i = 1}^{n} \left( \frac{\mathbb{I}\{\pi(X_i) = 1\} (Y_i - c)}{{e}(W_i = 1)} +\frac{\mathbb{I}\{\pi(X_i) = 0\} Y_i}{1 - {e}(W_i = 1)} \right)
\end{equation}
where $Y_i$ is a visit, and ${e}(W_i = 1) = 0.85$ is the probability of treatment assignment.

For each of the bootstrapped training samples, I estimate a targeting policy using a causal forest \citep{athey2019generalized}, a state-of-the-art machine learning method for estimating CATEs, which uses individual-level data of all 50,000 users and 12 covariates, and represents a benchmark of the targeting value with no privacy protections. I also estimate targeting policies under third-party privacy protections using uniform and strategic querying. For both of these querying methods, I collapse variables into three: $f0$ and $f6$, which are the two variables with variation across 25th, 50th, and 75th percentiles (see Table \ref{tab:criteo_summary}), are left as is, while all other variables are simply summed together into a third variable. On each sample, I impose four types of privacy protections: combinations of two levels of query limits (27 and 64)\footnote{For uniform querying, some of the covariate cells do not have enough (at least 20) people to be displayed. I skip these cells and reduce the number of queries available to strategic querying to match the number of actually performed uniform queries. This leads to the first query limit to be under 25, and the second limit to be under 50.} and two levels of differentially private noise (scale equal to 0.01, and 0.1).

The implementation of strategic querying has a few notable features. First, I use TAAF with no region size restrictions or penalty; however, I prevent the algorithm from sampling the exact same region it sampled before. Second, I estimate hyperparameters of the GP (amplitude, lengthscale, and noise) via MLE. I start the estimation at step 10, once I have collected enough data for the estimates to be stable.\footnote{Prior to step 10, I set fixed values of amplitude and lengthscale equal to 1 (variables are standardized) and noise scale equal to $s + 0.01$.} Third, I further encourage exploration by choosing the next region to query randomly among five regions with the highest values of TAAF and set $\beta$ in TAAF to $3 - \frac{i}{100}$, where $i$ is the query number.
\subsection{Results}

Table \ref{tab:targeting_results} compares  the targeting policies derived using the uniform querying and the strategic querying relative to the causal forest (no privacy protections) benchmark. I report the average lift over assigning everyone to treatment for each of the setting, as well as the average ratio of the lift relative to the causal forest lift. Since the results of the methods computed on bootstrapped samples are not independent, I first compute the ratio of the uniform and strategic querying methods relative to the causal forest and then average and compute confidence intervals across these ratios. 

\begin{table}[ht]
\centering
\caption{Comparison of targeting values across privacy settings}
\begin{tabular}{lcccc}
\toprule
& \makecell{\textbf{27 queries}\\\textbf{0.01 noise}} 
& \makecell{\textbf{27 queries}\\\textbf{0.1 noise}} 
& \makecell{\textbf{64 queries}\\\textbf{0.01 noise}} 
& \makecell{\textbf{64 queries}\\\textbf{0.1 noise}} \\
\midrule
\textbf{Causal Forest (mean)} & 17.87 & 17.87 & 17.87 & 17.87 \\
\addlinespace
\textbf{Uniform querying (mean)} & 16.81 & 14.82 & 13.72 & 6.28 \\
\quad Percentage of CF & 97\% & 85\% & 77\% & 33\% \\
\quad 95\% CI & (59\%, 148\%) & (33\%, 142\%) & (43\%, 117\%) & (--11\%, 80\%) \\
\addlinespace
\textbf{Strategic querying (mean)} & 17.15 & 17.43 & 17.78 & 17.64 \\
\quad Percentage of CF & 97\% & 100\% & 101\% & 101\% \\
\quad 95\% CI & (38\%, 151\%) & (50\%, 166\%) & (58\%, 152\%) & (49\%, 164\%) \\
\bottomrule
\end{tabular}

\justifying
\label{tab:targeting_results}
\small
\noindent
\textit{Note: Causal Forest is computed under no privacy protections and full access to the data. Percentages and confidence intervals are computed on the ratio of method performance relative to the causal forest baseline. Because values are computed from bootstrapped samples, the estimators are not independent, and the distribution of the ratio is used directly. The units of the outcomes are additional visits with a cost of treatment subtracted (see above).}
\end{table}

The uniform querying method performs very well with 27 queries and a low differential privacy noise, achieving 97\% of the non-privacy-preserving benchmark. However, when either the number of queries or the scale of differential privacy noise increases, its performance starts to suffer, dropping to just 33\% of the causal forest with 64 queries and 0.1 noise scale. While the detrimental effect of noise is expected, more queries should intuitively lead to better performance of the uniform querying method. There are, however, two mechanisms for a potential negative effect of more queries. First, similar to the example in Figure \ref{fig:uniform_problem}, the bins resulting from a smaller number of queries might better capture the relevant segments by chance if the treatment effect surface happened to align with the division of each variable by 3 rather than by 4. Second, in uniform querying, more queries result in smaller bins, which under differential privacy means that the signal-to-noise ratio is lower, and without smoothing and sharing information across cells, the uniform method cannot reliably detect the signs of the average values within each cell. Overall, these results suggest that the uniform querying method can be sufficient when (1) the marketer has a precise understanding of the expected granularity of the treatment effects and (2) the added differential privacy noise is low.

In contrast, the strategic querying method performs reliably well and is statistically indistinguishable from the causal forest across all four settings. Regardless of the number of queries and the scale of differential privacy noise, strategic querying achieves 97--101\% of the lift generated by the non-privacy-preserving causal forest. Compared to uniform querying, the strategic method dynamically adjusts the size of queries (they tend to decrease over time) as it learns the posterior of the treatment effects, and selects the regions to query that are the most consequential for targeting. Additionally, the inherent smoothing that the Gaussian Process prior imposes allows information sharing across cells and better distinction between noise inherent to the data and noise added through differential privacy. Because of these properties, the strategic querying method works well when the uniform does not: (1) when the granularity of the treatment effects is unknown or (2) when there are significant differential privacy protections.

The empirical results also more broadly challenge the conventional wisdom about targeting. Often, a targeting exercise becomes synonymous with precise estimation of individual-level heterogeneous treatment effects and thus antithetical to privacy protections. Yet, a targeting task is relatively ``easy'' compared to full estimation of HTEs, especially in realistic conditions when segments are big, as it only requires knowing the correct sign of the treatment effect for each segment. This can be accomplished with relatively light information requirements if done strategically as the results in Table \ref{tab:targeting_results} demonstrate: the strategic querying method that relies on fewer than 50\footnote{Even in the 64-query settings, the number of actual queries performed was under 50, see above.} noisy averages achieved the same targeting performance as a state-of-the-art machine learning method having access to 600,000 values.\footnote{50,000 rows of 12 variables.} Thus, targeting and privacy are not necessarily incompatible.

\section{Conclusion} \label{sec:conclusion}

Companies expect that in the next few years they will lose access to granular data and may only be able to access it through privacy-restricted APIs provided by platforms. In this paper, I developed a tool that allows marketers to design effective targeting policies even within restrictive privacy-preserving interfaces. The strategic querying method combines the ideas of Bayesian Optimization (dynamic exploration of a function space) with integral updating (querying regions instead of points) and a targeting-aware acquisition function (integrating the data exploration and the decision-making pipeline). In a simulation study, I showed that this method is expected to provide robust and effective targeting recommendations unless the conditions are extreme, such as a treatment effect surface that is extremely jagged, a small number of allowed queries, or differential privacy noise that is extremely high. In the empirical application over a large-scale dataset, strategic querying achieved 97--101\% of the performance of a state-of-the-art non-privacy-protected targeting method across all privacy protection settings, while a simpler uniform benchmark dropped to 33\% in some settings. Taken together, these results show that privacy protections and data-driven personalization are not necessarily mutually exclusive.

Future research can provide useful extensions of the framework developed in this paper. First, Bayesian Optimization can be combined with contextual bandits to aid variable selection and to handle the curse of dimensionality. Second, different privacy-preserving interfaces might provide information beyond group averages, such as variances, correlations, or linear regression coefficients. It would be useful to explore the best use and the value added of such information. Third, the framework presented in this paper has a convenient block-like structure, which allows it to be modified to fit marketers' needs. Knowledge of the data and past experiments can be seamlessly incorporated into the Gaussian Processes prior, and the acquisition function can be replaced to suit the task at hand. For example, future work can derive customized acquisition functions for segmentation, pricing, identifying the customers with the highest value, etc.

This paper also provides several managerial implications for both advertisers and advertising platforms. First, in the face of privacy protections, advertisers should focus on integrating their policy goals into data exploration pipelines and focus on obtaining information that is most consequential for their decisions rather than chase overall precision. Second, platforms can become more helpful for advertisers by incorporating elements of Bayesian predictive statistics into their interfaces and allowing marketers to construct custom acquisition functions which would guide data exploration. Finally, the results highlight that these two parties are not adversarial when it comes to data utility and privacy protections. Instead of negotiating over access to raw individual-level data, advertisers and platforms can collaborate to build better interfaces that would allow marketers to retain effectiveness in their decisions while protecting the privacy of platform users. 

\newpage
\setlength\bibsep{0pt}
\bibliographystyle{apalike}
\bibliography{literature}

\newpage
\appendix
\section{Appendix}

The three acquisition functions that I tested in a full simulation study (Section \ref{sec:simulation}) were chosen from a larger set of candidate functions that were pretested on 200 simulation runs. The other candidate functions include:

\begin{enumerate}
    \item \textbf{Pure variance.} $AF = \sigma$. Looks for regions with highest uncertainty, thereby learning the entire function with similar levels of precision \citep[e.g.,]{dew2024adaptive}.
    \item \textbf{Pure absolute mean.} $AF = |\mu|$. Looks for regions that are closest to zero in expectation, even if the uncertainty is low.
    \item \textbf{Log-weighted.} $AF = \ln \sigma^2 - \ln |\mu|$. Attempts to normalize the variance of large regions.
    \item \textbf{Area-weighted.} $AF = \frac{\sigma - \mu}{area}$. Attempts to penalize large regions. 
    \item \textbf{Ratio.} $AF = \frac{\sigma}{|\mu|}$. An alternative functional form of TAAF.  
    \item \textbf{Full posterior update.} For each of the candidate regions, values are sampled from a posterior predictive distribution. Within each sample, the treatment effects function is reestimated using the sampled value and the expected value of the targeting policy is recalculated. Such values are then averaged over samples for each candidate region. The region with the highest expected average improvement in targeting policy is queried next.
\end{enumerate}

\end{document}